\documentclass[%
aip,
pop,%
amsmath,amssymb,
reprint,%
]{revtex4-1}

\usepackage{graphicx}% Include figure files
\usepackage{epspdfconversion}
\usepackage{dcolumn}% Align table columns on decimal point
\usepackage{bm}% bold math
\begin{document}

\preprint{AIP/123-QED}

\title{Memory effects in the velocity relaxation process of the dust particle\\ in dusty plasma}% Force line breaks with \\

\author{Z. Ghannad}
\email{z.ghannad2020@gmail.com}%Lines break automatically or can be forced with \\
\author{H. Hakimi Pajouh}%
\email{hakimipajouh@gmail.com}
\affiliation{%
Faculty of Physics and Chemistry, Alzahra University, P. O. Box 19938-93973, Tehran, Iran
}%


\begin{abstract}
In this paper, by comparing the time scales associated with the velocity  relaxation and correlation time of the random
force due to dust charge fluctuations, memory effects in the velocity relaxation of an isolated dust particle exposed to the random force due to dust charge fluctuations are considered, and the velocity relaxation process of the dust particle is considered as a non-Markovian stochastic process. Considering memory effects in the velocity relaxation process of the dust particle yields a retarded friction force, which is introduced by a memory kernel in the fractional Langevin equation. The fluctuation-dissipation theorem for the dust grain is derived from this equation. The mean-square displacement and the velocity autocorrelation function of the
dust particle are obtained, and their asymptotic behavior, the
dust particle temperature due to charge 
fluctuations, and the diffusion coefficient are studied in
the long-time limit. As an interesting feature, it is found that by considering memory effects in the velocity relaxation process of the dust particle, fluctuating force on the dust particle can cause an anomalous diffusion in a dusty plasma. In this case, the
mean-square displacement of the dust grain increases slower than linearly with time, and the velocity
autocorrelation function decays as a power-law instead of the exponential decay. Finally, in the Markov limit, these results are in good agreement with those obtained from previous works for Markov (memoryless) process of the velocity relaxation.
\end{abstract}

\maketitle

%\tableofcontents

\section{\label{sec:level1}Introduction}

 Dust particles in a dusty plasma acquire a net electric charge
by collecting electrons and ions from the background plasma.
The dust grain charge fluctuates in time because of the discrete nature of charge carriers~\cite{Kampen}. Electrons and ions arrive at the dust surface at random times. For this reason, the charge fluctuates. These fluctuations always exist even in a steady-state uniform plasma~\cite{Cui}.

Dust charge fluctuations have been investigated by many researchers~\cite{Matsoukas1996,Matsoukas1997,Matsoukas1995,Shotorban2014,Khrapak,Shotorban2011,Matthews}. In a dusty plasma, there are many phenomena that dust charge fluctuations can be considered as a reason for them such as heating of dust particles system~\cite{Vaulina,Nefedov,Angelis}, instability of lattice oscillations in a low-pressure gas discharge~\cite{Ivlev2000,Morfill}, and the formation of the shock waves in dusty plasmas~\cite{Mamun}. Also, the motion of dust particles under the influence of the random force due to dust charge fluctuations has been investigated in some studies~\cite{Ivlev2010,Quinn,Schmidt}. In these studies, the motion of dust particles has been modeled by Brownian motion based on the Fokker-Planck or Langevin equations, assuming that dust charge fluctuations are fast. It means that the relaxation timescale for charge fluctuations is much shorter than the relaxation time for the dust velocity; hence, they have considered the stochastic motion of dust particles as a Markov process with no memory. It means that the stochastic motion of the dust after the time $t$ is entirely independent of its history before the time $t$, i.e., the dust particle has no memory of the past.

The values of the relaxation timescale for dust charge fluctuations and the dust velocity entirely depend on the dusty plasma parameters. For example, Hoang $\textit{et al.}$~\cite{Hoang} showed that the relaxation time of dust charge fluctuations $\tau_c$ is comparable to the relaxation time of the dust velocity $\tau_r$, i.e., $\tau_c\approx\tau_r$ for very small dust particles with radius $a\leqslant 5\times10^{-8}$ cm in the interstellar medium. In addition to space dusty plasmas, in laboratory dusty plasmas depending on the plasma parameters, such as pressure or density of the neutral gas, $\tau_c$ is comparable to $\tau_r$. As a result, in such situations, the main assumptions of fast charge fluctuations and the Markov process for the velocity relaxation of the dust particle become inappropriate. Thus, memory effects are important in the velocity relaxation of dust particles as a non-Markov process, and they cannot generally be neglected.

In this paper, we study memory effects in the velocity relaxation of the dust particle exposed to the random force due to charge fluctuations. We present an analytic model based on a fractional Langevin equation. We will show that in the presence of memory effects in the velocity relaxation, dust charge fluctuations can cause an anomalous diffusion of the dust particle in a dusty plasma. The anomalous diffusion of dust particles has been experimentally observed in laboratory dusty plasmas~\cite{Nunomura,Juan}, and our research provides a possible reason for this behavior based on the memory effects. It is important to note here that the diffusion of a dust particle means a process of random displacements of a dust particle in a specified time interval.

The paper is organized as follows. In section \ref{Timescales}, we introduce major timescales characterizing
dynamics of the dust particle. In section \ref{model}, a model based on the fractional Langevin equation is presented and solved using the Laplace transform technique. In section \ref{calculation}, we calculate the mean-square displacement and the velocity autocorrelation function of the dust grain. Section \ref{limit} is devoted to the analysis of the asymptotic behavior of the results. Section \ref{summary}
 contains summary and conclusions.

\section{\label{Timescales}Timescales}
Let us consider an isolated spherical dust particle in the sheath, and study the relaxation timescales of dust charge fluctuations and the dust velocity. The particle is charged by collecting electrons and ions from the plasma. The particle charge fluctuates about the steady-state value because of the discrete nature of the electron and ion currents~\cite{Cui}.
 It was shown that the autocorrelation function of dust charge fluctuations has the following form~\cite{Khrapak}
\begin{equation}
\langle \delta Z(t)\delta Z(t')\rangle=\langle \delta Z^2\rangle\, \rm{exp}\left(-\beta\vert t-t'\vert\right),\label{eq1}
\end{equation}
where $\delta Z(t)$=$Z(t)$--$Z_0$, $Z(t)$ is the instantaneous charge, $Z_0$ is the steady-state (equilibrium) charge, $\langle \delta Z^2\rangle$ is the square of the amplitude of random charge fluctuations, and charging frequency $\beta$ is defined as the relaxation frequency for small deviations of the charge from the equilibrium value $Z_0$. For the isolated dust particle under the condition $a\ll\lambda_D\ll\lambda_{\mathrm{mfp}}$, where $a$ is the dust radius, $\lambda_D$ is the screening length due to electrons and ions, and $\lambda_{\mathrm{mfp}}$ is the mean free path for electron-neutral or ion-neutral collisions, the charging frequency can be calculated by using  the orbital- motion-limited (OML) theory~\cite{Fortov2010}
\begin{equation}
\beta=-\frac{d(I_i-I_e)}{dZ}\bigg|_{Z=Z_0}=\frac{1+z}{\sqrt{2\pi}}\frac{a}{\lambda_{Di}}\omega_{pi}\label{eq2}
\end{equation}
where $I_i$=$\sqrt{8\pi}a^2n_iv_{Ti}\left(1-Ze^2/aT_i\right)$ is the ion flux, $I_e$=$\sqrt{8\pi}a^2n_ev_{Te} \rm{exp}$$\left(Ze^2/aT_e\right)$ is the electron flux to the particle surface, $v_{Ti(e)}$=$\left(T_{i(e)}/m_{i(e)}\right)^{1/2}$ is the ion (electron) thermal velocity, $T_{i(e)}$, $m_{i(e)}$, and $n_{i(e)}$ are ion (electron) temperature, mass, and number density, respectively, $z$=$\vert Z \vert e^2/aT_e$ is the absolute magnitude of the particle charge in the units $aT_e/e^2$, $\lambda_{Di}$=$\sqrt{T_i/4\pi e^2n_i}$ is the ionic Debye radius, and $\omega_{pi}$=$v_{Ti}/\lambda_{Di}$ is the ion plasma frequency.

Let us now consider an isolated dust particle with fluctuating charge in the sheath. To separate the dust particle transport due to dust charge fluctuations from other processes, which in turn can influence the dust motion in the plasma sheath, we only consider the  gravitational, electric field, and neutral drag forces on the dust. We hence neglect the ion drag, electron drag, and thermophoretic forces, collisions between dust particles, and other processes, which can be included in more realistic models~\cite{Khrapak2}. The motion of this dust particle is treated as a stochastic process because of the stochastic nature of the force due to dust charge fluctuations, and it can generally be modeled by a normal Langevin equation of the form~\cite{Vaulina} 
\begin{equation}
\dot{v}(t)+\gamma v(t)=f(t)+\xi(t),\label{eq3}
\end{equation}
where $v(t)$ is the velocity of the dust particle, $-\gamma v(t)$ is the neutral drag force per unit mass,$\gamma$ is the damping
rate due to neutral gas friction, $\xi(t)$ is the stochastic Langevin force per unit mass due to collisions with the neutral gas molecules, and $f(t)$ is the random force per unit mass representing the effect of dust charge fluctuations. Here, the forces acting on a dust are the electric force due to the sheath electric field, the gravitational force, and the stochastic Langevin force, i.e., $F(t)$=$F_Z(t)$+$F_g$+$\xi(t)$. The electric force is given by $F_Z(t)$=$eEZ(t)$, where $Z(t)$=$Z_0$+$\delta Z(t)$ is the instantaneous dust charge, and $E$ is the electric field. Note that, for simplicity, we have neglected the fluctuations of the electric field. The force can be written as $F(t)$=$F_0$+$f(t)$+$\xi(t)$, where $F_0$=$eEZ_0$+$F_g$. Thus, the random force due to dust charge fluctuations has the following form
\begin{equation}
f(t)=eE\delta Z(t).\label{eq4}
\end{equation}
In a steady state $eEZ_0$+$F_g$=$0$, so that $F(t)$=$f(t)$+$\xi(t)$. By using Eqs. (\ref{eq4}) and~(\ref{eq1}), one can see that the random force due to dust charge fluctuations per unit mass has following properties
\begin{equation}
\langle f(t)\rangle=0,\,\,\langle f(t)f(t')\rangle=\frac{e^2E^2}{m_d^2}\langle \delta Z^2\rangle\, \rm{exp}\left(-\beta\vert \it{t-t'}\vert\right)\label{eq5}
\end{equation}
where $m_d$ is the dust particle mass.
The stochastic Langevin force per unit mass has following properties
\begin{equation}
\langle \xi(t)\rangle=0,\,\,\,\,\,\,\langle \xi(t)\xi(t')\rangle=S_n\delta(t-t'),\label{eq6}
\end{equation}
where $S_n$ is the intensity of the stochastic Langevin force.
By solving the Eq.~(\ref{eq3}), the dust velocity is obtained in the following form
\begin{equation}
v(t)=v(0)e^{-\gamma t}+\int_0^t \left(\it{f(t')+\xi(t')}\right)e^{-\gamma(t-t')} dt';\label{eq7}
\end{equation}
then, mean-square velocity is obtained in the form of
\begin{equation*}
\langle v^2(t)\rangle=v(0)e^{-2\gamma t}+\int_0^t\int_0^tdt'dt'' e^{-\gamma(t-t')} e^{-\gamma(t-t'')}\times
\end{equation*}
\begin{equation}
\,\,\,\,\,\,\,\,\,\,\,\,\,\,\,\,\,\,\,\,\,\,\,\,\,\langle\left(f(t')+\xi(t')\right)\left(f(t'')+\xi(t'')\right)\rangle.\label{eq8}
\end{equation}
Then, by using the fact that the stochastic Langevin force and the random force due to dust charge fluctuations come from different sources; therefore, they are uncorrelated and independent, so that 
\begin{equation*}
\langle f(t')\xi(t'')\rangle=\langle\xi(t') f(t'')\rangle=0,
\end{equation*}
then by substituting in Eq.~(\ref{eq8}), and using Eqs.~(\ref{eq5}) and~(\ref{eq6}), we obtain
\begin{equation*}
\langle v^2(t)\rangle=v(0)e^{-2\gamma t}+\int_0^t\int_0^tdt'dt'' e^{-\gamma(t-t')} e^{-\gamma(t-t'')}\times
\end{equation*}
\begin{equation}
\left(S_n\delta(t'-t'')+\frac{e^2E^2}{m_d^2}\langle \delta Z^2\rangle e^{-\beta(t'-t'')}\right).\label{eq9}
\end{equation}
By taking the integral and applying the limit $t\to\infty$, we obtain the long-time behavior of the mean-square velocity in the following form
\begin{equation}
\langle v^2(t)\rangle_{t\to\infty}=\frac{S_n}{2\gamma}+\frac{e^2E^2}{2m_d^2\beta\gamma}\langle \delta Z^2\rangle.\label{eq10}
\end{equation}
The mean-square velocity in long-time limit is representative of the dust temperature
\begin{equation}
T_d=m_d\langle v^2(t)\rangle_{t\to\infty}\label{eq11}
\end{equation}
therefore, from Eq.~(\ref{eq10}), one can see that the dust temperature has two part. One part is due to collisions with the neutral gas molecules, $T_n$, and another part is due to dust charge fluctuations, $T_f$, so that
\begin{equation}
T_d=T_n+T_f,\label{eq12}
\end{equation}
where 
\begin{equation*}
T_n=\frac{m_dS_n}{2\gamma},\,\,\,\,\,\,\,T_f=\frac{e^2E^2}{2m_d\beta\gamma}\langle \delta Z^2\rangle.
\end{equation*}
Note that since we want to study the role of dust charge fluctuations in the dust particle transport, we neglect the stochastic Langevin force, and we assume that the random force due to dust charge fluctuations is important. Therefore, in the absence of stochastic Langevin force ($\xi(t)$=0), $T_d$=$T_f$, and the Langevin equation Eq.~(\ref{eq3}) reduces to
\begin{equation}
\dot{v}(t)+\gamma v(t)=f(t),\label{eq13}
\end{equation}

Now, we study the major timescales characterizing dynamics of the charged grain. First, we introduce a dimensionless parameter $\epsilon$, which characterizes memory effects in the velocity relaxation for stochastic processes
\begin{equation}
\epsilon=\frac{\tau_r}{\tau_c},\label{eq14}
\end{equation}
\begin{equation}
\tau_r=\frac{1}{C_v(0)}\int_0^\infty C_v(t)dt,\,\,\,\,\tau_c=\frac{1}{C_f(0)}\int_0^\infty C_f(t)dt\label{eq15}
\end{equation}
where $\tau_r$ is the relaxation time of the velocity, $\tau_c$ is the correlation (relaxation) time of the random force, $C_v(t)=\langle v(0)v(t)\rangle$ is the velocity autocorrelation function (VACF), and $C_f(t)$=$\langle f(0)f(t) \rangle$ is the random force autocorrelation function. In general, stochastic processes are classified into two types of Markov and non-Markov processes  based on the timescales~\cite{Ridolfi,Mori}:
\begin{itemize}
\item A stochastic process is said to have memory effects in the velocity relaxation, if its relaxation time of the velocity is comparable to the relaxation time of the random force, i.e., $\tau_r\approx\tau_c$. Then, according to Eq.~(\ref{eq14}), we find that $\epsilon\approx 1$. In this situation, the stochastic process is called a non-Markov process with memory effects in the velocity relaxation. The memory in the velocity means that the velocity of the particle at the current time depends on its velocity at all past times.
\item However, the situation $\tau_r\gg\tau_c$ corresponds to a memoryless behavior, meaning that the velocity at the current time is entirely independent of the velocity at all past times. In this case, the stochastic process is called a Markov process, and according to Eq.~(\ref{eq14}), we find that $\epsilon\to\infty$.
\end{itemize}

Let us calculate the relaxation timescales for the dust velocity and the random force. We obtain the dust velocity from Eq.~(\ref{eq13}) in the following form
\begin{equation}
v(t)=v(0)e^{-\gamma t}+\int_0^t \it{f(t')}e^{-\gamma(t-t')} dt'.\label{eq16}
\end{equation}
Multiplying Eq.~(\ref{eq16}) by $v(0)$, and performing an appropriate ensemble average $\langle...\rangle$, and by using $\langle v(0)f(t)\rangle=0$, we obtain
\begin{equation}
C_v(t)=\langle v(0)^2\rangle e^{-\gamma t};\label{eq17}
\end{equation}
then, by substituting Eq.~(\ref{eq17}) into~(\ref{eq15}), we find
\begin{equation}
\tau_r=\frac{1}{\langle v(0)^2\rangle}\int_0^\infty \langle v(0)^2\rangle e^{-\gamma t}dt=\frac{1}{\gamma}.\label{eq18}
\end{equation}
The random force autocorrelation function is obtained by substituting $t'=0$ into Eq.~(\ref{eq5})
\begin{equation*}
C_f(t)=\frac{e^2E^2}{m_d^2}\langle \delta Z^2\rangle\, \rm{exp}\left(\it{-\beta t}\right),
\end{equation*}
then, by substituting $C_f(t)$ into Eq.~(\ref{eq15}), one finds
\begin{equation}
\tau_c=\int_0^\infty e^{-\it{\beta t}}dt=\frac{1}{\beta}.\label{eq19}
\end{equation}
When the damping rate of the neutral gas is much smaller than the dust charging frequency, i.e., $\gamma\ll\beta$, the stochastic process for the velocity $v(t)$ of the dust particle can be considered as a Markov process, and memory effects in the velovity relaxation can reasonably be neglected. In this case, the Langevin equation (Eq.~(\ref{eq13})) is appropriate for the description of the dust motion under the influence of the random force due to charge fluctuations. In general, the values of the damping rate and the charging frequency entirely depend on the plasma parameters. For example, we consider three different types of plasmas with neon, argon, and krypton  neutral gases, and estimate the values of $\gamma$ and $\beta$ using typical experimental values for various parameters. We assume  the neutral gases at room temperature and in the range of pressures 0.5--1.0 Torr. We also assume $T_e$=4 $\rm{eV}, \it{T_e/T_i}$=40, $n_i$=$10^8 \rm{cm^{-3}}$, and the silica dust particle with radius $a$=0.5 $\mu$m and the mass density $\rho$=2 $\rm{g/cm^3}$. For neon, argon, and krypton neutral gases with $T_e/T_i$=40, the absolute magnitudes of the dust charge are $z\sim$2.6, 2.8, and 3.2, respectively~\cite{Fortov2010,Fortov2004}. With the given parameters, we calculate the charging frequency from Eq.~(\ref{eq2}). The values of the relaxation time of the random force due to dust charge fluctuations $\tau_c$=$\beta^{-1}$ (in seconds) are listed in the last column of Table~\ref{jlab1}.

To calculate the timescale $\tau_r$ from Eq.~(\ref{eq18}), we first need to find the damping rate. When the neutral gas mean free path is long compared to the dust grain radius, it is appropriate to use the Epstein drag force to calculate $\gamma$~\cite{Epstein,Baines}. The mean free path values of neon, argon, and krypton atoms at the maximal
pressure used (1.0 Torr) are approximately equal to 92, 60, and 40 $\mu$m, respectively. These are about 184, 120, and 80 times larger than the dust grain radius (0.5 $\mu$m). Thus, the damping rate is given by the Epstein formula
\begin{equation}
\gamma=\delta\sqrt{\frac{8}{\pi}}\left(\frac{T_g}{m_g}\right)^{-\frac{1}{2}}\frac{P}{\rho a},\label{eq20}
\end{equation}
where $m_g$, $T_g$, $P$ are the mass, temperature, and pressure of the neutral gas, respectively. The parameter $\delta$ is 1 for specular reflection or 1.39 for diffuse reflection of the neutral gas atoms from the dust grain~\cite{Epstein,Baines}. We use $\delta$=1.39 and the given parameters to calculate the damping rate from Eq.~(\ref{eq20}). The relaxation time values of the dust velocity 
$\tau_r$=$\gamma^{-1}$ (in seconds) are tabulated in Table~\ref{jlab1} for each gas at various pressures. As shown in Table~\ref{jlab1}, 
with increasing the pressure (or equivalently increasing the density) of the neutral gas, the relaxation time of the dust velocity becomes comparable to the relaxation time of the random force. Thus, in this situation, the assumption of the fast rate for charge fluctuations becomes inappropriate, and memory effects in the velocity of the dust particle cannot be neglected. As a result, the normal Langevin equation becomes inappropriate for the description of the dust motion because this equation is built on the Markovian assumption with no memory. When  $\tau_r$ becomes comparable to $\tau_c$, the memory effects in the relaxation of the dust velocity become important, because at the times of the same order of  $\tau_c$, the random forces due to dust charge fluctuations are correlated. As a result, the velocity of the dust particle at the current time depends on its velocity at past times, and it means that the process of the dust velocity relaxation is retarded, and these retarded effects (or equivalently memory effects) are characterized with memory kernel in the friction force within Langevin equation. Now, this equation with retarded friction is called the fractional Langevin equation. Hence, the fractional Langevin equation is built on the non-Markovian assumption, while the Langevin equation without the existence a retarded friction is built on the Markov assumption. Note that, as shown in Table~\ref{jlab1}, we study the regime $\tau_r\approx\tau_c$ not $\tau_r\ll\tau_c$, because in the regime  $\tau_r\ll\tau_c$, the damping rate of the neutral gas is very high and the kinetic energy transferred to the dust (due to charge fluctuations) is totally dissipated by friction with the neutral gas, as mentioned in Ref. [10].

 In the next section, we introduce a model based on fractional Langevin equation for the evolution process of the dust particle as a non-Markov process with memory effects in the velocity.
\begin{table}
\caption{\label{jlab1}The values of the relaxation time of the dust velocity $\tau_r$ for each gas at various pressures. $P$ in Torr, $\tau_r$ and $\tau_c$ in s. With increasing the gas pressure, $\tau_r$ becomes comparable to $\tau_c$. }
\begin{ruledtabular}
\bgroup
\def\arraystretch{1.6}%  1 is the default, change whatever you need
\begin{tabular}{cccccccc}
Gas &$P$=0.5&$P$=0.8&$P$=1.0&
 &$\tau_c$ \\
\hline
Kr&1.2$\times$$10^{-3}$&7.3$\times$$10^{-4}$&5.8$\times$$10^{-4}$&
& 1.9$\times$$10^{-4}$\\
Ar&1.7$\times$$10^{-3}$&1.1$\times$$10^{-3}$&8.4$\times$$10^{-4}$ &
&1.4$\times$$10^{-4}$\\
Ne&2.4$\times$$10^{-3}$ &1.5$\times$$10^{-3}$&1.2$\times$$10^{-3}$ &
&1.1$\times$$10^{-4}$\\
\end{tabular}
\egroup
\end{ruledtabular}
\end{table}

\section{\label{model}Basic equation and fluctuation-dissipation theorem}
Now, we consider memory effects in the velocity relaxation of an isolated dust  particle exposed to the random force due to dust charge fluctuations. We model the motion of the dust grain based on the fractional Langevin equation (FLE) because this equation includes a retarded friction force with a memory kernel function, which is non-local in time, and shows memory effects in the velocity relaxation of the dust particle. The FLE is as follows~\cite{Burov}
\begin{equation}
\dot{v}(t)+\frac{\bar{\gamma}}{\Gamma(1-{\alpha})}\int_0^t \left(\frac{\vert t-t'\vert}{\tau_c}\right)^{-\alpha}v(t')dt'=f(t),\label{eq21}
\end{equation}
where 0$<\alpha<$1, and $\Gamma(1-{\alpha})$ is the gamma function. $\bar{\gamma}$ is the scaling factor with physical dimension (time)$^{-2}$, and must be introduced to ensure the correct dimension of the equation. We define $\bar{\gamma}=(\tau_r\tau_c)^{-1}$, so that for $\alpha$=1, and according to the Dirac generalized function, $\delta(t-t')$=$\vert t-t'\vert^{-1}/\Gamma(0)$~\cite{Gel'fand}, Eq.~(\ref{eq21}) reduces to the normal Langevin equation (13).

Equation~(\ref{eq21}) can be rewritten in the following form
\begin{equation}
\dot{v}(t)+\int_0^t\gamma(t-t')v(t')dt'=f(t),\label{eq22}
\end{equation}
where
\begin{equation}
\gamma(t-t')=\frac{(\tau_r\tau_c)^{-1}}{\Gamma(1-\alpha)}\left(\frac{\vert t-t'\vert}{\tau_c}\right)^{-\alpha}\label{eq23}
\end{equation}
 is often called the memory kernel function.
It is interesting to know that the name fractional in the fractional Langevin equation originates from the fractional derivative, which is defined in the Caputo sense as follows~\cite{Caputo}
\begin{equation*}
\frac{d^{\alpha}f(t)}{dt^{\alpha}}={_0}{\mathcal{D}}{_t^{\alpha-1}}\left(\frac{df(t)}{dt}\right),
\end{equation*}
where ${_0}{\mathcal{D}}{_t^{\alpha-1}}$ is the Riemann-Liouville fractional integral~\cite{Miller,Podlubny}
\begin{equation}
{_0}{\mathcal{D}}{_t^{\alpha-1}}f(t)=\frac{1}{\Gamma(1-\alpha)}\int_0^t\left(t-t'\right)^{-\alpha}f(t')dt';\label{eq24}
\end{equation}
consequently, the fractional Langevin equation reads
\begin{equation*}
\dot{v}(t)+\frac{\bar{\gamma}}{\tau_c^{-\alpha}}\frac{d^{\alpha}x(t)}{dt^{\alpha}}=f(t);
\end{equation*}
therefore, the name fractional Langevin equation is confirmed.

Now, we need to derive a suitable relation between the memory kernel and the autocorrelation function of the random force. To this end, we first define the Fourier transforms for the velocity and random force as follows
\begin{equation}
v(\omega)=\int_{-\infty}^{\infty}v(t)\,e^{i\omega t}dt,\,\,\,\,f(\omega)=\int_{-\infty}^{\infty}f(t)\,e^{i\omega t}dt.\label{eq25}
\end{equation}
Using Eq.~(\ref{eq25}) and taking the Fourier transform of Eq.~(\ref{eq22}) yields
\begin{equation}
v(\omega)=\frac{f(\omega)}{-i\omega+\gamma(\omega)}\label{eq26}
\end{equation}
where $\gamma(\omega)$, defined by $\gamma(\omega)=\int_0^\infty \gamma(t)\,e^{i\omega t}dt$,
denotes the Fourier-Laplace transform of the memory kernel $\gamma(t)$. The velocity autocorrelation function is defined by the formula:
\begin{equation}
C_v(\tau)=\langle v(t)v(t+\tau)\rangle=\lim_{\theta \rightarrow \infty}\frac{1}{\theta}\int_{-\theta/2}^{\theta/2}v(t)v(t+\tau)dt,\label{eq27}
\end{equation}
where $\theta$ is the time interval for the integration. Generally, the power spectrum $S(\omega)$ and the autocorrelation function of a stochastic process $C(\tau)$ are related by the Wiener-Khintchine theorem \cite{Yates}
\begin{subequations}
\begin{equation}
S(\omega)=\int_{-\infty}^{\infty}C(\tau)\,e^{i\omega \tau}d\tau,\label{subeq:1}
\end{equation}
\begin{equation}
C(\tau)=\frac{1}{2\pi}\int_{-\infty}^{\infty}S(\omega)\,e^{-i\omega \tau}d\omega.\label{subeq:2}
\end{equation}
\end{subequations}
By substituting $C_v(\tau)$ from Eq.~(\ref{eq27}) into (\ref{subeq:1}) and using Eq.~(\ref{eq25}), we obtain the velocity power spectrum as
\begin{equation}
S_v(\omega)=\lim_{\theta \rightarrow \infty}\frac{1}{\theta}\vert v(\omega)\vert^2;\label{eq29}
\end{equation}
then, by using Eq.~(\ref{eq26}), we obtain the relation between the power spectrums of the velocity and random force
\begin{equation}
S_v(\omega)=\frac{S_f(\omega)}{\vert\gamma(\omega)-i\omega\vert^2},\label{eq30}
\end{equation}
where $S_f(\omega)$=$\lim_{\theta \rightarrow \infty}\frac{1}{\theta}\vert f(\omega)\vert^2$ is the power spectrum of the random force. Note that Eq.~(\ref{eq30}) was obtained by using the fractional Langevin equation for the non-Markov velocity process.
In the same way, we obtain the relation between the power spectrums of the velocity and random force for the Markov velocity process by using the normal Langevin equation in the following form
\begin{equation}
S_v(\omega)=\frac{S_f(\omega)}{\vert\gamma-i\omega\vert^2}.\label{eq31}
\end{equation}
When the power spectrum of the random force $S_f(\omega)$ is given, the above equation yields the velocity power spectrum $S_v(\omega)$ from which the VACF is obtained by Eq.~(\ref{subeq:2}). If VACF should include the velocity in the thermal equilibrium (i.e., at sufficiently long times), $S_f(\omega)$ is required to satisfy a special condition. Now, we obtain this condition. In the long-time limit $(\vert t-t'\vert\gg\tau_c)$, the autocorrelation function of the random force obtained from Eq.~(\ref{eq5}) reduces to the following form
\begin{equation}
C_f(\tau)=\langle f(t)f(t+\tau)\rangle=\frac{e^2E^2}{m_d^2\beta}\langle \delta Z^2\rangle \delta(\tau);\label{eq32}
\end{equation}
then, by using Eq.~(\ref{subeq:1}), the force power spectrum reads
\begin{equation}
S_f=\frac{e^2E^2}{m_d^2\beta}\langle \delta Z^2\rangle.\label{eq33}
\end{equation}
The mean-square velocity $\langle v^2\rangle$ can be evaluated by substituting $\tau$=0 into Eq.~(\ref{eq27}). Thus, by using Eqs.~(\ref{eq33}),~(\ref{eq31}), and~(\ref{subeq:2}), one finds
\begin{equation}
\langle v^2\rangle=\frac{S_f}{2\gamma}=\frac{e^2E^2\langle\delta Z^2\rangle}{2m_d^2\beta\gamma}.\label{eq34}
\end{equation}
As we mentioned before, the mean-square velocity in the long-time limit (the equilibrium velocity) is representative of the dust temperature
\begin{equation}
T_d=m_d\langle v^2\rangle.\label{eq35}
\end{equation}
Thus, by using Eqs. (\ref{eq34}) and (\ref{eq35}), the special condition for the power spectrum of the random force $S_f(\omega)$ is obtained by
\begin{equation}
S_f=\frac{2T_d\gamma}{m_d}.\label{eq36}
\end{equation}
It is important to note that this condition, i.e., Eq.~(\ref{eq36}), was obtained by using Eq.~(\ref{eq31}) for the Markov velocity process. For the non-Markov velocity process given by the FLE , the relation between the power spectrums of the velocity and random force is given by Eq.~(\ref{eq30}) instead of~(\ref{eq31}); hence, the condition for the power spectrum $S_f(\omega)$, in this case, is a generalization of Eq.~(\ref{eq36}) as follows
\begin{equation}
S_f(\omega)=\frac{2T_d}{m_d}\rm{Re}\boldsymbol{\left(}\gamma(\omega)\boldsymbol{\right)};\label{eq37}
\end{equation}
then, by using Eqs. (\ref{eq25}), (\ref{subeq:1}), and (\ref{eq37}), we obtain
\begin{equation*}
\langle f(\omega)f^*(\omega')\rangle=2\pi S_f(\omega)\delta(\omega-\omega')
\end{equation*}
\begin{equation}
\langle f(\omega)f^*(\omega')\rangle=\frac{4\pi T_d}{m_d}\rm{Re}\boldsymbol{\left(}\gamma(\omega)\delta(\omega-\omega')\boldsymbol{\right)},\label{eq38}
\end{equation}
where $f^*(\omega')$ is the complex conjugate of the function $f(\omega')$. Therefore, we obtain the relation between the autocorrelation function of the random force and the memory kernel by using the inverse Fourier transform of  Eq.~(\ref{eq38})
\begin{equation}
\langle f(t)f(t')\rangle=\frac{T_d}{m_d}\gamma(t-t')=\frac{S_f}{2\gamma}\gamma(t-t').\label{eq39}
\end{equation}
We call this the fluctuation-dissipation theorem for the dust particle because dust charge fluctuations ($S_f\propto\langle \delta Z^2\rangle$, according to Eq.~(\ref{eq33})) are necessarily accompanied by the friction ($\gamma$ in the denominator). It is important to note that in FLE, we have normalized the time $\vert t-t'\vert$ to the correlation time $\tau_c$. The reason for this can be found from the fluctuation-dissipation theorem. According to Eq.~(\ref{eq23}), for $\vert t-t'\vert\gg\tau_c$, the memory kernel relaxes to zero. Consequently, by using the fluctuation-dissipation theorem, $\langle f(t)f(t')\rangle$ relaxes to zero. Thus, the characteristic time of the memory relaxation is the correlation time of the random force due to charge fluctuations.

Now, we solve Eq.~(\ref{eq22}) for the dust velocity with the initial conditions $x_0$=$x(0)$ and $v_0$=$v(0)$, by using the Laplace transform technique. We obtain
\begin{equation}
v(t)=\langle v(t)\rangle+\int_0^t g(t-t')f(t')dt',\label{eq40}
\end{equation}
where $\langle v(t)\rangle$=$v_0g(t)$ is the mean dust velocity, and the function $g(t)$ is the inverse Laplace transform of
\begin{equation}
\hat{g}(s)=\frac{1}{s+\hat{\gamma}(s)}=\frac{1}{s+\gamma\left(\tau_c s\right)^{\alpha -1}},\label{eq41}
\end{equation}
where $\hat{\gamma}(s)$=$\gamma\left(\tau_c s\right)^{\alpha -1}$ is the Laplace transform of the memory kernel $\gamma(t)$, and the Laplace transform of the function $f(t)$ is defined by $\hat{f}(s)$=$\int_0^\infty f(t)e^{-\it{st}}dt$.
To find the function $g(t)$, we first need to introduce the Mittag-Leffler (ML) function $E_{\alpha}\left(at^{\alpha}\right)$ as follows \cite{Erdelyi}
\begin{equation}
E_{\alpha}\left(at^{\alpha}\right)=\sum_{n=0}^{\infty}\frac{\left(at^{\alpha}\right)^n}{\Gamma(\alpha n+1)},\,\,\,\,\,\alpha>0.\label{eq42}
\end{equation}
The ML function is a generalization of the exponential function. For $\alpha$=1, we have the exponential function, $E_1(at)$=$e^{\it{at}}$. The Laplace transform of the ML function is given by \cite{Haubold}
\begin{equation}
\int_0^\infty E_{\alpha}(at^{\alpha})e^{-\it{st}}dt=\frac{s^{-1}}{1-as^{-\alpha}}.\label{eq43}
\end{equation}
Thus, by using Eqs. (\ref{eq41}) and (\ref{eq43}), one finds
\begin{equation}
g(t)=E_{2-\alpha}\boldsymbol{\left(}-\gamma\tau_c\left(t/\tau_c\right)^{2-\alpha}\boldsymbol{\right)}.\label{eq44}
\end{equation}

The dust particle velocity is the derivative of the dust position. Therefore, the position of the dust particle is obtained by 
\begin{equation}
x(t)=\langle x(t)\rangle+\int_0^t G(t-t')f(t')dt',\label{eq45}
\end{equation}
where $\langle x(t)\rangle$=$x_0+v_0G(t)$ is the mean dust particle position, and $G(t)$=$\int_0^t g(t')dt'$. To calculate the function $G(t)$ in terms of the ML function, we use the following identity \cite{Haubold}
\begin{equation}
\frac{d}{dt}t^{\beta-1}E_{\alpha,\beta}\left(at^{\alpha}\right)=t^{\beta-2}E_{\alpha,\beta-1}\left(at^{\alpha}\right),\label{eq46}
\end{equation}
where 
\begin{equation}
E_{\alpha,\beta}\left(at^{\alpha}\right)=\sum_{n=0}^\infty\frac{\left(at^{\alpha}\right)^n}{\Gamma(\alpha n+\beta)},\,\,\,\,\,\alpha,\beta>0\label{eq47}
\end{equation}
is the generalization of the ML function that for $\beta$=1 reduces to Eq.~(\ref{eq42}), i.e., $E_{\alpha,1}\left(at^{\alpha}\right)$=$E_{\alpha}\left(at^{\alpha}\right)$ \cite{Shukla}. Then, by using Eqs. (\ref{eq44}), (\ref{eq46}), and also using $g(t)$=$dG(t)/dt$, one reads
\begin{equation}
G(t)=tE_{2-\alpha,2}\boldsymbol{\left(}-\gamma\tau_c\left(t/\tau_c\right)^{2-\alpha}\boldsymbol{\right)}.\label{eq48}
\end{equation}
It is worth mentioning that the functions $G(t)$ and $g(t)$ help us find the mean-square displacement (MSD) and the velocity autocorrelation function of the dust particle. In the next section, we evaluate the MSD and VACF for the dust particle.

\section{\label{calculation}Mean-square displacement and velocity autocorrelation function}
Below, we calculate two diagnostics to characterize the random motion of the dust particle due to random charge fluctuations. The first diagnostic is MSD, and it is the most common quantitative tool used to
investigate random processes. The MSD of a dust particle is defined by the relation
\begin{equation}
MSD=\langle (x(t)-x_0)^2 \rangle=(\langle x(t) \rangle-x_0)^2+\sigma_x^2,\label{eq49}
\end{equation}
where $\sigma_x^2$=$\langle x^2(t) \rangle- \langle x(t) \rangle^2$ is the variance of the displacement, and $\langle...\rangle$ denotes an
average over an ensemble of random trajectories. By using Eqs. (\ref{eq45}), (\ref{eq39}), and (\ref{eq34}), we have
\begin{eqnarray}
\sigma_x^2=\frac{e^2E^2\langle\delta Z^2\rangle}{2m_d^2\beta\gamma}& \int_0^t dt'\int_0^tdt''G(t-t')\nonumber\\
&\times G(t-t'')\gamma(t'-t'').\label{eq50}
\end{eqnarray}
Using the double Laplace transform technique \cite{Vinales,Pottier2003}, and after some calculations, we obtain the following expression for the variance
\begin{equation}
\sigma_x^2=\frac{e^2E^2\langle\delta Z^2\rangle}{2m_d^2\beta\gamma}\left(2I(t)-G^2(t)\right),\label{eq51}
\end{equation}
where $I(t)=\int_0^tG(t')dt'$. The function $I(t)$ is obtained from Eqs. (\ref{eq46}) and (\ref{eq48}) as follows
\begin{equation}
I(t)=t^2E_{2-\alpha,3}\boldsymbol{\left(}-\gamma\tau_c\left(t/\tau_c\right)^{2-\alpha}\boldsymbol{\right)}.\label{eq52}
\end{equation}
By substituting Eq.~(\ref{eq51}) into~(\ref{eq49}), and using $x_0=0$ and $\langle x(t) \rangle=v_0G(t)$, we have
\begin{equation*}
MSD=G^2(t)\left(v_0^2-\frac{e^2E^2\langle\delta Z^2\rangle}{2m_d^2\beta\gamma}\right)
\end{equation*}
\begin{equation}
+\frac{e^2E^2\langle\delta Z^2\rangle}{m_d^2\beta\gamma}I(t);\label{eq53}
\end{equation}
then, from Eqs. (\ref{eq48}) and (\ref{eq52}), we get
\begin{equation*}
MSD=\frac{e^2E^2\langle\delta Z^2\rangle}{m_d^2\beta\gamma}t^2 E_{2-\alpha,3}\boldsymbol{\left(}-\gamma\tau_c\left(t/\tau_c\right)^{2-\alpha}\boldsymbol{\right)}
\end{equation*}
\begin{equation}
\,\,\,\,\,\,\,\,\,\,\,+t^2E^2_{2-\alpha,2}\boldsymbol{\left(}-\gamma\tau_c\left(t/\tau_c\right)^{2-\alpha}\boldsymbol{\right)}\left(v_0^2-\frac{e^2E^2\langle\delta Z^2\rangle}{2m_d^2\beta\gamma}\right).\label{eq54}
\end{equation}

The second diagnostic, VACF, is the average of the initial velocity of a dust particle multiplied by its velocity at a later time. To calculate the VACF of the dust grain, we first calculate the double Laplace transform $\langle \hat{v}(s)\hat{v}(s')\rangle$ of the function $\langle \hat{v}(t)\hat{v}(t')\rangle$, and we obtain
\begin{equation*}
\langle \hat{v}(s)\hat{v}(s')\rangle=\left(v_0^2-\frac{e^2E^2\langle\delta Z^2\rangle}{2m_d^2\beta\gamma}\right)\hat{g}(s)\hat{g}(s')
\end{equation*}
\begin{equation}
\,\,\,\,\,\,\,\,\,\,\,\,\,+\,\frac{e^2E^2\langle\delta Z^2\rangle}{2m_d^2\beta\gamma}\frac{\hat{g}(s)+\hat{g}(s')}{s+s'},\label{eq55}
\end{equation}
where $\mathit{\hat{v}}(s)$, $\mathit{\hat{g}}(s)$, and $\mathit{\hat{G}}(s)$ are Laplace transforms of $\mathit{v}(t)$, 
$\mathit{g}(t)$, and $G(t)$, respectively. Taking the inverse double Laplace transform of Eq.~(\ref{eq55}) yields
\begin{equation*}
\langle v(t)v(t')\rangle=\left(v_0^2-\frac{e^2E^2\langle\delta Z^2\rangle}{2m_d^2\beta\gamma}\right)g(t)g(t')
\end{equation*}
\begin{equation}
\,\,\,\,\,\,+\,\frac{e^2E^2\langle\delta Z^2\rangle}{2m_d^2\beta\gamma}g(t-t');\label{eq56}
\end{equation}
therefore, by substituting $t'$=0 into Eq.~(\ref{eq56}), the VACF can be written as
\begin{equation}
C_v(t)=v_0^2g(t)=v_0^2E_{2-\alpha}\boldsymbol{\left(}-\gamma\tau_c\left(t/\tau_c\right)^{2-\alpha}\boldsymbol{\right)}.\label{eq57}
\end{equation}
By substituting $\alpha$=1 into Eq.~(\ref{eq57}), we obtain
\begin{equation}
C_v(t)=v_0^2 e^{-\gamma t},\label{eq58}
\end{equation}
which corresponds to the function $C_v(t)$ in Eq.~(\ref{eq17}) obtained by the Langevin equation (13).
As we expected, the velocity autocorrelation function of dust particle decays exponentially with time for the memoryless case.

\section{\label{limit}Asymptotic behaviors}
In order to gain physical insight, we investigate the long-time behavior of the MSD and VACF. We first study the asymptotic behavior of the function $g(t)$ for $t\gg\tau_c$.
 The asymptotic behavior of the Mittag-Leffler functions is given by \cite{Erdelyi}
\begin{equation}
E_{\alpha}(y)\sim-\frac{y^{-1}}{\Gamma(1-\alpha)},\,\,\,y\to\infty,\label{eq59}
\end{equation}
\begin{equation}
E_{\alpha,\beta}(y)\sim-\frac{y^{-1}}{\Gamma(\beta-\alpha)},\,\,\,y\to\infty.\label{eq60}
\end{equation}
By using Eqs. (\ref{eq59}) and (\ref{eq44}), we have
\begin{equation}
g(t)\sim\frac{1}{\gamma\tau_c\Gamma(\alpha-1)}\left(\frac{t}{\tau_c}\right)^{\alpha-2}.\label{eq61}
\end{equation}
Also, by using Eqs. (\ref{eq60}), (\ref{eq48}), and (\ref{eq52}), we read
\begin{equation}
G(t)\sim\frac{1}{\gamma\Gamma(\alpha)}\left(\frac{t}{\tau_c}\right)^{\alpha-1},\label{eq62}
\end{equation}
and 
\begin{equation}
I(t)\sim\frac{\tau_c}{\gamma\Gamma(1+\alpha)}\left(\frac{t}{\tau_c}\right)^{\alpha}.\label{eq63}
\end{equation}

To calculate the dust temperature due to charge fluctuations, by substituting $t'$=$t$ into Eq.~(\ref{eq56}); then, by using Eq.~(\ref{eq61}), we obtain
\begin{equation*}
\langle v^2(t)\rangle\sim\left(v_0^2-\frac{e^2E^2\langle\delta Z^2\rangle}{2m_d^2\beta\gamma}\right)\frac{(t/\tau_c)^{2\alpha-4}}{\left(\gamma\tau_c\Gamma(\alpha-1)\right)^2}
\end{equation*}
\begin{equation}
+\,\frac{e^2E^2\langle\delta Z^2\rangle}{2m_d^2\beta\gamma}.\label{eq64}
\end{equation}
We observe that a slow power-law decay like $t^{2\alpha-4}$ for the mean-square velocity. In the long-time limit, by applying $t\gg\tau_c$ to Eq.~(\ref{eq64}), we obtain the temperature of the dust particle as follows
\begin{equation}
T_d=m_d\langle v^2\rangle=\frac{e^2E^2\langle\delta Z^2\rangle}{2m_d\beta\gamma},\label{eq65}
\end{equation}
which coincides with the temperature obtained from the Markov velocity process~\cite{Vaulina}.

\begin{figure}[t]
\begin{center}
\includegraphics[width=9cm,height=6.75cm]{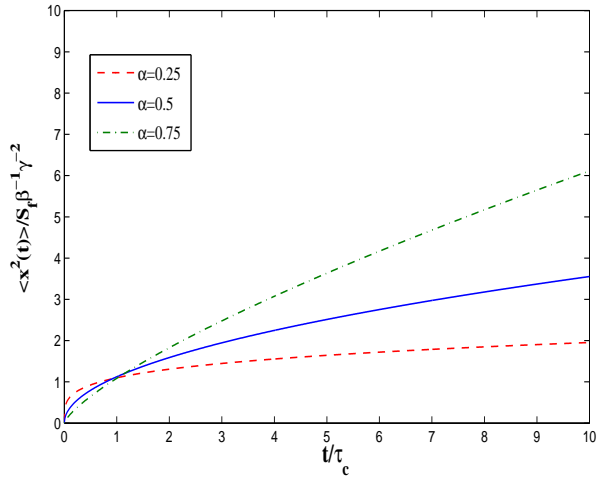}
\caption{\label{figure1} Normalized mean-square displacements as a function of time: anomalous diffusion for $\alpha$=0.25, 0.5, 0.75 (dashed line, solid line, dashed-dotted line).  }
\end{center}
\end{figure}

To find the long-time behavior of the MSD and VACF of the dust particle, without loss of generality, we can choose the initial conditions in the following form
\begin{equation}
x_0=0,\,\,\,\,v_0^2=\frac{e^2E^2\langle\delta Z^2\rangle}{2m_d^2\beta\gamma},\label{eq66}
\end{equation}
where $v_0^2$ is the mean-square velocity in the thermal equilibrium given by Eq.~(\ref{eq34}).
Thus, by using Eqs. (\ref{eq53}), (\ref{eq66}), and (\ref{eq63}), MSD can be written as
\begin{equation}
MSD=\langle x^2(t) \rangle\sim\frac{e^2E^2\langle\delta Z^2\rangle}{m_d^2\beta^2\gamma^2\Gamma(1+\alpha)}
\left(\frac{t}{\tau_c}\right)^{\alpha}.\label{eq67}
\end{equation}
The MSD obtained from Eq.~(\ref{eq67}) is a nonlinear function of time because the exponent $\alpha$ takes the values $0<\alpha<1$. It implies that fluctuating force by considering memory effects in the velocity relaxation of the dust particle leads to the anomalous diffusion of the dust particle. We emphasize here that the diffusion of a dust particle means a process of random displacements of a dust particle in a specified time interval.
In Figure~\ref{figure1}, we display the curves $\langle x^2(t) \rangle$ for the values $\alpha$=0.25, 0.5, and 0.75. For the given parameters $\alpha$, the MSD increases asymptotically slower than linearly with time (the signature of anomalous diffusion).

Diffusion coefficient for the anomalous diffusion processes is given by the generalized Green-Kubo relation in the following form~\cite{Kneller}
\begin{equation}
D_{\alpha}=\frac{1}{\Gamma({1+\alpha})}\int_0^\infty {_0}{\mathcal{D}}{_t^{\alpha-1}}C_v(t)dt,\label{eq68}
\end{equation}
where $D_{\alpha}$ is often called the generalized diffusion coefficient, and ${_0}{\mathcal{D}}{_t^{\alpha-1}}$ is the Riemann-Liouville fractional integral [see Eq.~(\ref{eq24})] . For $\alpha$=1, this relation reduces to the standard Green-Kubo relation, which holds for the normal diffusion~\cite{Green,Kubo}
\begin{equation*}
D=\int_0^\infty C_v(t)dt.
\end{equation*}
Generalized diffusion coefficient can easily be obtained by the Laplace transform of Eq.~(\ref{eq68}). The Laplace transform of the Riemann-Liouville fractional integral is given by~\cite{Podlubny}
\begin{equation}
\int_0^\infty {_0}{\mathcal{D}}{_t^{\alpha-1}} C_v(t)e^{-st}dt=s^{\alpha-1}\hat{C}_v(s);\label{eq69}
\end{equation}
then, by using 
\begin{equation*}
\lim_{s \rightarrow 0}\int_0^\infty {_0}{\mathcal{D}}{_t^{\alpha-1}} C_v(t)e^{-st}dt=\int_0^\infty {_0}{\mathcal{D}}{_t^{\alpha-1}} C_v(t)dt,
\end{equation*}
and using Eq.~(\ref{eq68}), we find
\begin{equation}
D_{\alpha}=\lim_{s \rightarrow 0}\frac{1}{\Gamma(1+\alpha)}s^{\alpha-1}\hat{C}_v(s).\label{eq70}
\end{equation}
The function $\hat{C}_v(s)$ is obtained by using Eqs.~(\ref{eq57}),~(\ref{eq41}), and~(\ref{eq66}) as follows
\begin{equation}
\hat{C}_v(s)=v_0^2\hat{g}(s)=\frac{e^2E^2\langle\delta Z^2\rangle}{2m_d^2\beta\gamma}\frac{1}{s+\gamma(\tau_c s)^{\alpha-1}};\label{eq71}
\end{equation}
therefore, by substituting Eq.~(\ref{eq71}) into~(\ref{eq70}), we find the generalized diffusion coefficient of the dust particle
\begin{equation}
D_{\alpha}=\frac{e^2E^2\langle\delta Z^2\rangle}{2\Gamma(1+\alpha)m_d^2\beta^2\gamma^2\tau_c^{\alpha}},\label{eq72}
\end{equation}
which has a physical dimension (length)$^2$/(time)$^{\alpha}$.

\begin{figure}
\begin{center}
\includegraphics[width=9cm,height=6.75cm]{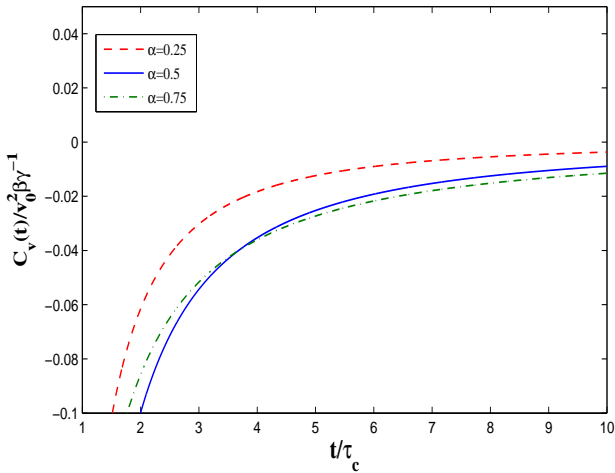}
\caption{\label{figure2} Normalized velocity autocorrelation functions as a function of time for $\alpha$=0.25, 0.5, 0.75 (dashed line, solid line, dashed-dotted line). The Curves of $C_v(t)$ are negative at all times.}
\end{center}
\end{figure}

Moreover, using Eqs.~(\ref{eq57}),~(\ref{eq61}), and~(\ref{eq66}), the long-time behavior of the velocity autocorrelation function of the dust particle can be written as
\begin{equation}
C_v(t)\sim\frac{e^2E^2\langle\delta Z^2\rangle}{2m_d^2\gamma^2\Gamma(\alpha-1)}\left(\frac{t}{\tau_c}\right)^{\alpha-2}.\label{eq73}
\end{equation}
Thus, the function $C_v(t)$ for long times exhibits a power-law decay instead of the exponential decay [see Eq.~(\ref{eq17})]. The power-law decay can obviously be seen in Figure~\ref{figure2} that we show the curves $C_v(t)$ for the values $\alpha$=0.25, 0.5, and 0.75. For the given values of the parameter $\alpha$ , the function $\Gamma(\alpha-1)$ is negative, and the curves $C_v(t)$ have a negative tail at all times, $C_v(t)<0$. As we mentioned,  the function $C_v(t)$ is the average of the velocity of a dust particle at the time $t$ multiplied by its velocity at a later time. When $C_v(t)$ is negative, this means that there are anti-correlations between the dust particle velocities at the time $t$ and the later time $t+t'$, i.e., the diffusing dust particle tends to change the direction of its motion and goes back, which indicates an anti-persistent motion for the dust particle. In other words, the motion of the dust particle in a direction at the time $t$ will be followed by its motion in the opposite direction at the later time, and the dust particle prefers to continually change its direction instead of continuing in the same direction; as a result, the diffusion of the dust grain is slower than normal case in normal diffusion.

As we expected, in the very long times limit, the velocity autocorrelation function of the dust particle decays to zero. It can also be seen by applying the limit $t\to\infty$ to Eq.~(\ref{eq73}), and this result is consistent with the result $C_v(t)\to0$ obtained from Markov dynamics (by applying the limit $t\to\infty$ to $C_v(t)$=$v_0^2e^{-\gamma t}$ from Eq.~(\ref{eq17}), $C_v(t)$ goes to zero). The reason for this consistency between the results (i.e., vanishing $C_v(t)$ in the very long times limit) obtained from non-Markov and Markov dynamics can be understood as follows. As we showed, at the times of the order of $\tau_r\approx\tau_c$, memory effects in the velocity relaxation process of the dust particle become important, and their effects yield to power-law behavior for MSD and VACF. However, in very long times, i.e., when the observation time $t$ is much longer than all characteristic time scales including the correlation time of the random force due to dust charge fluctuations $\tau_c$ and the relaxation time of the dust velocity $\tau_r$, there are no longer the memory effects in the velocity of the dust particle, and the dynamics of the dust particle in the very long times limit is a Markov dynamics with no memory effects in the velocity. Therefore, as in Markov dynamics in the limit $t\to\infty$, the function $C_v(t)$ goes to zero, the function $C_v(t)$ obtained from Eq.~(\ref{eq73}) in the limit $t\to\infty$ approaches to zero.
\section{\label{summary}Summary and conclusions}
We have studied memory effects in the velocity relaxation of an isolated dust particle. 
First, we have compared the relaxation timescales of the fluctuating force and the dust velocity, and have shown that they can be of the same order of the magnitude, depending on the plasma parameters. 
Thus, the fast charge fluctuations assumption does not always hold, and memory effects in the velocity relaxation should generally be considered. 
We have developed a model based on the fractional Langevin equation for the evolution of the dust particle. Memory effects in the velocity relaxation of the dust particle have been introduced by using the memory kernel in this equation. 
We have derived a suitable fluctuation-dissipation theorem for the dust grain, which relates the autocorrelation function of the random force to the memory kernel.
Then, the fractional Langevin equation has been solved by the Laplace transform technique, and the mean-square displacement and the velocity autocorrelation function of the dust grain have been obtained in terms of the generalized Mittag-Leffler functions. 
We have investigated the asymptotic behaviors of the MSD, VACF, generalized diffusion coefficient, and the dust temperature due to charge fluctuations in the long-time limit. We have found that in the presence of memory effects in the relaxation of the dust velocity, dust charge fluctuations can cause the anomalous diffusion of the dust particle, which has been experimentally observed in laboratory dusty plasmas~\cite{Nunomura,Juan}.
In this case, the mean-square displacement of the dust grain has a nonlinear dependence on time, and the velocity autocorrelation function decays as a power-law instead of the exponential decay.

\nocite{*}

\bibliography{Ghannad}% Produces the bibliography via BibTeX.

\providecommand{\noopsort}[1]{}\providecommand{\singleletter}[1]{#1}%
\begin{thebibliography}{42}%
\makeatletter
\providecommand \@ifxundefined [1]{%
 \@ifx{#1\undefined}
}%
\providecommand \@ifnum [1]{%
 \ifnum #1\expandafter \@firstoftwo
 \else \expandafter \@secondoftwo
 \fi
}%
\providecommand \@ifx [1]{%
 \ifx #1\expandafter \@firstoftwo
 \else \expandafter \@secondoftwo
 \fi
}%
\providecommand \natexlab [1]{#1}%
\providecommand \enquote  [1]{``#1''}%
\providecommand \bibnamefont  [1]{#1}%
\providecommand \bibfnamefont [1]{#1}%
\providecommand \citenamefont [1]{#1}%
\providecommand \href@noop [0]{\@secondoftwo}%
\providecommand \href [0]{\begingroup \@sanitize@url \@href}%
\providecommand \@href[1]{\@@startlink{#1}\@@href}%
\providecommand \@@href[1]{\endgroup#1\@@endlink}%
\providecommand \@sanitize@url [0]{\catcode `\\12\catcode `\$12\catcode
  `\&12\catcode `\#12\catcode `\^12\catcode `\_12\catcode `\%12\relax}%
\providecommand \@@startlink[1]{}%
\providecommand \@@endlink[0]{}%
\providecommand \url  [0]{\begingroup\@sanitize@url \@url }%
\providecommand \@url [1]{\endgroup\@href {#1}{\urlprefix }}%
\providecommand \urlprefix  [0]{URL }%
\providecommand \Eprint [0]{\href }%
\providecommand \doibase [0]{http://dx.doi.org/}%
\providecommand \selectlanguage [0]{\@gobble}%
\providecommand \bibinfo  [0]{\@secondoftwo}%
\providecommand \bibfield  [0]{\@secondoftwo}%
\providecommand \translation [1]{[#1]}%
\providecommand \BibitemOpen [0]{}%
\providecommand \bibitemStop [0]{}%
\providecommand \bibitemNoStop [0]{.\EOS\space}%
\providecommand \EOS [0]{\spacefactor3000\relax}%
\providecommand \BibitemShut  [1]{\csname bibitem#1\endcsname}%
\let\auto@bib@innerbib\@empty
%</preamble>
\bibitem [{\citenamefont {van Kampen}(2007)}]{Kampen}%
  \BibitemOpen
  \bibfield  {author} {\bibinfo {author} {\bibfnamefont {N.~G.}\ \bibnamefont
  {van Kampen}},\ }\href@noop {} {\emph {\bibinfo {title} {Stochastic Processes
  in Physics and Chemistry}}}\ (\bibinfo  {publisher} {Elsevier, Amsterdam},\
  \bibinfo {year} {2007})\BibitemShut {NoStop}%
\bibitem [{\citenamefont {Cui}\ and\ \citenamefont {Goree}(1994)}]{Cui}%
  \BibitemOpen
  \bibfield  {author} {\bibinfo {author} {\bibfnamefont {C.}~\bibnamefont
  {Cui}}\ and\ \bibinfo {author} {\bibfnamefont {J.}~\bibnamefont {Goree}},\
  }\href@noop {} {\bibfield  {journal} {\bibinfo  {journal} {IEEE Trans. Plasma
  Sci}\ }\textbf {\bibinfo {volume} {22}},\ \bibinfo {pages} {151} (\bibinfo
  {year} {1994})}\BibitemShut {NoStop}%
\bibitem [{\citenamefont {Matsoukas}, \citenamefont {Russell},\ and\
  \citenamefont {Smith}(1996)}]{Matsoukas1996}%
  \BibitemOpen
  \bibfield  {author} {\bibinfo {author} {\bibfnamefont {T.}~\bibnamefont
  {Matsoukas}}, \bibinfo {author} {\bibfnamefont {M.}~\bibnamefont {Russell}},
  \ and\ \bibinfo {author} {\bibfnamefont {M.}~\bibnamefont {Smith}},\
  }\href@noop {} {\bibfield  {journal} {\bibinfo  {journal} {J. Vac. Sci.
  Technol. A}\ }\textbf {\bibinfo {volume} {14}},\ \bibinfo {pages} {624}
  (\bibinfo {year} {1996})}\BibitemShut {NoStop}%
\bibitem [{\citenamefont {Matsoukas}\ and\ \citenamefont
  {Russell}(1997)}]{Matsoukas1997}%
  \BibitemOpen
  \bibfield  {author} {\bibinfo {author} {\bibfnamefont {T.}~\bibnamefont
  {Matsoukas}}\ and\ \bibinfo {author} {\bibfnamefont {M.}~\bibnamefont
  {Russell}},\ }\href@noop {} {\bibfield  {journal} {\bibinfo  {journal} {Phys.
  Rev. E}\ }\textbf {\bibinfo {volume} {55}},\ \bibinfo {pages} {991} (\bibinfo
  {year} {1997})}\BibitemShut {NoStop}%
\bibitem [{\citenamefont {Matsoukas}\ and\ \citenamefont
  {Russell}(1995)}]{Matsoukas1995}%
  \BibitemOpen
  \bibfield  {author} {\bibinfo {author} {\bibfnamefont {T.}~\bibnamefont
  {Matsoukas}}\ and\ \bibinfo {author} {\bibfnamefont {M.}~\bibnamefont
  {Russell}},\ }\href@noop {} {\bibfield  {journal} {\bibinfo  {journal} {J.
  Appl. Phys.}\ }\textbf {\bibinfo {volume} {77}},\ \bibinfo {pages} {4285}
  (\bibinfo {year} {1995})}\BibitemShut {NoStop}%
\bibitem [{\citenamefont {Shotorban}(2014)}]{Shotorban2014}%
  \BibitemOpen
  \bibfield  {author} {\bibinfo {author} {\bibfnamefont {B.}~\bibnamefont
  {Shotorban}},\ }\href@noop {} {\bibfield  {journal} {\bibinfo  {journal}
  {Phys. Plasmas}\ }\textbf {\bibinfo {volume} {21}},\ \bibinfo {pages}
  {033702} (\bibinfo {year} {2014})}\BibitemShut {NoStop}%
\bibitem [{\citenamefont {Khrapak}\ \emph {et~al.}(1999)\citenamefont
  {Khrapak}, \citenamefont {Nefedov}, \citenamefont {Petrov},\ and\
  \citenamefont {Vaulina}}]{Khrapak}%
  \BibitemOpen
  \bibfield  {author} {\bibinfo {author} {\bibfnamefont {S.~A.}\ \bibnamefont
  {Khrapak}}, \bibinfo {author} {\bibfnamefont {A.~P.}\ \bibnamefont
  {Nefedov}}, \bibinfo {author} {\bibfnamefont {O.~F.}\ \bibnamefont {Petrov}},
  \ and\ \bibinfo {author} {\bibfnamefont {O.~S.}\ \bibnamefont {Vaulina}},\
  }\href@noop {} {\bibfield  {journal} {\bibinfo  {journal} {Phys. Rev. E}\
  }\textbf {\bibinfo {volume} {59}},\ \bibinfo {pages} {6017} (\bibinfo {year}
  {1999})}\BibitemShut {NoStop}%
\bibitem [{\citenamefont {Shotorban}(2011)}]{Shotorban2011}%
  \BibitemOpen
  \bibfield  {author} {\bibinfo {author} {\bibfnamefont {B.}~\bibnamefont
  {Shotorban}},\ }\href@noop {} {\bibfield  {journal} {\bibinfo  {journal}
  {Phys. Rev. E}\ }\textbf {\bibinfo {volume} {83}},\ \bibinfo {pages} {066403}
  (\bibinfo {year} {2011})}\BibitemShut {NoStop}%
\bibitem [{\citenamefont {Matthews}, \citenamefont {Shotorban},\ and\
  \citenamefont {Hyde}(2013)}]{Matthews}%
  \BibitemOpen
  \bibfield  {author} {\bibinfo {author} {\bibfnamefont {L.~S.}\ \bibnamefont
  {Matthews}}, \bibinfo {author} {\bibfnamefont {B.}~\bibnamefont {Shotorban}},
  \ and\ \bibinfo {author} {\bibfnamefont {T.~W.}\ \bibnamefont {Hyde}},\
  }\href@noop {} {\bibfield  {journal} {\bibinfo  {journal} {Astrophys. J}\
  }\textbf {\bibinfo {volume} {776}},\ \bibinfo {pages} {103} (\bibinfo {year}
  {2013})}\BibitemShut {NoStop}%
\bibitem [{\citenamefont {Vaulina}\ \emph
  {et~al.}(1999{\natexlab{a}})\citenamefont {Vaulina}, \citenamefont {Khrapak},
  \citenamefont {Nefedov},\ and\ \citenamefont {Petrov}}]{Vaulina}%
  \BibitemOpen
  \bibfield  {author} {\bibinfo {author} {\bibfnamefont {O.~S.}\ \bibnamefont
  {Vaulina}}, \bibinfo {author} {\bibfnamefont {S.~A.}\ \bibnamefont
  {Khrapak}}, \bibinfo {author} {\bibfnamefont {A.~P.}\ \bibnamefont
  {Nefedov}}, \ and\ \bibinfo {author} {\bibfnamefont {O.~F.}\ \bibnamefont
  {Petrov}},\ }\href@noop {} {\bibfield  {journal} {\bibinfo  {journal} {Phys.
  Rev. E}\ }\textbf {\bibinfo {volume} {60}},\ \bibinfo {pages} {5959}
  (\bibinfo {year} {1999}{\natexlab{a}})}\BibitemShut {NoStop}%
\bibitem [{\citenamefont {Vaulina}\ \emph
  {et~al.}(1999{\natexlab{b}})\citenamefont {Vaulina}, \citenamefont {Nefedov},
  \citenamefont {Petrov},\ and\ \citenamefont {Khrapak}}]{Nefedov}%
  \BibitemOpen
  \bibfield  {author} {\bibinfo {author} {\bibfnamefont {O.~S.}\ \bibnamefont
  {Vaulina}}, \bibinfo {author} {\bibfnamefont {A.~P.}\ \bibnamefont
  {Nefedov}}, \bibinfo {author} {\bibfnamefont {O.~F.}\ \bibnamefont {Petrov}},
  \ and\ \bibinfo {author} {\bibfnamefont {S.~A.}\ \bibnamefont {Khrapak}},\
  }\href@noop {} {\bibfield  {journal} {\bibinfo  {journal} {J. Exp. Theor.
  Phys.}\ }\textbf {\bibinfo {volume} {88}},\ \bibinfo {pages} {1130} (\bibinfo
  {year} {1999}{\natexlab{b}})}\BibitemShut {NoStop}%
\bibitem [{\citenamefont {de~Angelis}\ \emph {et~al.}(2005)\citenamefont
  {de~Angelis}, \citenamefont {Ivlev}, \citenamefont {Morfill},\ and\
  \citenamefont {Tsytovich}}]{Angelis}%
  \BibitemOpen
  \bibfield  {author} {\bibinfo {author} {\bibfnamefont {U.}~\bibnamefont
  {de~Angelis}}, \bibinfo {author} {\bibfnamefont {A.~V.}\ \bibnamefont
  {Ivlev}}, \bibinfo {author} {\bibfnamefont {G.~E.}\ \bibnamefont {Morfill}},
  \ and\ \bibinfo {author} {\bibfnamefont {V.~N.}\ \bibnamefont {Tsytovich}},\
  }\href@noop {} {\bibfield  {journal} {\bibinfo  {journal} {Phys. Plasmas}\
  }\textbf {\bibinfo {volume} {12}},\ \bibinfo {pages} {052301} (\bibinfo
  {year} {2005})}\BibitemShut {NoStop}%
\bibitem [{\citenamefont {Ivlev}, \citenamefont {Konopka},\ and\ \citenamefont
  {Morfill}(2000)}]{Ivlev2000}%
  \BibitemOpen
  \bibfield  {author} {\bibinfo {author} {\bibfnamefont {A.~V.}\ \bibnamefont
  {Ivlev}}, \bibinfo {author} {\bibfnamefont {U.}~\bibnamefont {Konopka}}, \
  and\ \bibinfo {author} {\bibfnamefont {G.}~\bibnamefont {Morfill}},\
  }\href@noop {} {\bibfield  {journal} {\bibinfo  {journal} {Phys. Rev. E}\
  }\textbf {\bibinfo {volume} {62}},\ \bibinfo {pages} {2739} (\bibinfo {year}
  {2000})}\BibitemShut {NoStop}%
\bibitem [{\citenamefont {Morfill}, \citenamefont {Ivlev},\ and\ \citenamefont
  {Jokipii}(1999)}]{Morfill}%
  \BibitemOpen
  \bibfield  {author} {\bibinfo {author} {\bibfnamefont {G.}~\bibnamefont
  {Morfill}}, \bibinfo {author} {\bibfnamefont {A.~V.}\ \bibnamefont {Ivlev}},
  \ and\ \bibinfo {author} {\bibfnamefont {J.~R.}\ \bibnamefont {Jokipii}},\
  }\href@noop {} {\bibfield  {journal} {\bibinfo  {journal} {Phys. Rev. Lett.}\
  }\textbf {\bibinfo {volume} {83}},\ \bibinfo {pages} {971} (\bibinfo {year}
  {1999})}\BibitemShut {NoStop}%
\bibitem [{\citenamefont {Mamun}\ and\ \citenamefont {Shukla}(2002)}]{Mamun}%
  \BibitemOpen
  \bibfield  {author} {\bibinfo {author} {\bibfnamefont {A.~A.}\ \bibnamefont
  {Mamun}}\ and\ \bibinfo {author} {\bibfnamefont {P.~K.}\ \bibnamefont
  {Shukla}},\ }\href@noop {} {\bibfield  {journal} {\bibinfo  {journal} {IEEE
  Trans. Plasma Sci.}\ }\textbf {\bibinfo {volume} {30}},\ \bibinfo {pages}
  {720} (\bibinfo {year} {2002})}\BibitemShut {NoStop}%
\bibitem [{\citenamefont {Ivlev}\ \emph {et~al.}(2010)\citenamefont {Ivlev},
  \citenamefont {Lazarian}, \citenamefont {Tsytovich}, \citenamefont
  {de~Angelis}, \citenamefont {Hoang},\ and\ \citenamefont
  {Morfill}}]{Ivlev2010}%
  \BibitemOpen
  \bibfield  {author} {\bibinfo {author} {\bibfnamefont {A.~V.}\ \bibnamefont
  {Ivlev}}, \bibinfo {author} {\bibfnamefont {A.}~\bibnamefont {Lazarian}},
  \bibinfo {author} {\bibfnamefont {V.~N.}\ \bibnamefont {Tsytovich}}, \bibinfo
  {author} {\bibfnamefont {U.}~\bibnamefont {de~Angelis}}, \bibinfo {author}
  {\bibfnamefont {T.}~\bibnamefont {Hoang}}, \ and\ \bibinfo {author}
  {\bibfnamefont {G.~E.}\ \bibnamefont {Morfill}},\ }\href@noop {} {\bibfield
  {journal} {\bibinfo  {journal} {Astrophys. J}\ }\textbf {\bibinfo {volume}
  {723}},\ \bibinfo {pages} {612} (\bibinfo {year} {2010})}\BibitemShut
  {NoStop}%
\bibitem [{\citenamefont {Quinn}\ and\ \citenamefont {Goree}(2000)}]{Quinn}%
  \BibitemOpen
  \bibfield  {author} {\bibinfo {author} {\bibfnamefont {R.~A.}\ \bibnamefont
  {Quinn}}\ and\ \bibinfo {author} {\bibfnamefont {J.}~\bibnamefont {Goree}},\
  }\href@noop {} {\bibfield  {journal} {\bibinfo  {journal} {Phys. Rev. E}\
  }\textbf {\bibinfo {volume} {61}},\ \bibinfo {pages} {3033} (\bibinfo {year}
  {2000})}\BibitemShut {NoStop}%
\bibitem [{\citenamefont {Schmidt}\ and\ \citenamefont {Piel}(2015)}]{Schmidt}%
  \BibitemOpen
  \bibfield  {author} {\bibinfo {author} {\bibfnamefont {C.}~\bibnamefont
  {Schmidt}}\ and\ \bibinfo {author} {\bibfnamefont {A.}~\bibnamefont {Piel}},\
  }\href@noop {} {\bibfield  {journal} {\bibinfo  {journal} {Phys. Rev. E}\
  }\textbf {\bibinfo {volume} {92}},\ \bibinfo {pages} {043106} (\bibinfo
  {year} {2015})}\BibitemShut {NoStop}%
\bibitem [{\citenamefont {Hoang}\ and\ \citenamefont {Lazarian}(2012)}]{Hoang}%
  \BibitemOpen
  \bibfield  {author} {\bibinfo {author} {\bibfnamefont {T.}~\bibnamefont
  {Hoang}}\ and\ \bibinfo {author} {\bibfnamefont {A.}~\bibnamefont
  {Lazarian}},\ }\href@noop {} {\bibfield  {journal} {\bibinfo  {journal}
  {Astrophys. J}\ }\textbf {\bibinfo {volume} {761}},\ \bibinfo {pages} {96}
  (\bibinfo {year} {2012})}\BibitemShut {NoStop}%
\bibitem [{\citenamefont {Nunomura}\ \emph {et~al.}(2006)\citenamefont
  {Nunomura}, \citenamefont {Samsonov}, \citenamefont {Zhdanov},\ and\
  \citenamefont {Morfill}}]{Nunomura}%
  \BibitemOpen
  \bibfield  {author} {\bibinfo {author} {\bibfnamefont {S.}~\bibnamefont
  {Nunomura}}, \bibinfo {author} {\bibfnamefont {D.}~\bibnamefont {Samsonov}},
  \bibinfo {author} {\bibfnamefont {S.}~\bibnamefont {Zhdanov}}, \ and\
  \bibinfo {author} {\bibfnamefont {G.}~\bibnamefont {Morfill}},\ }\href@noop
  {} {\bibfield  {journal} {\bibinfo  {journal} {Phys. Rev. Lett.}\ }\textbf
  {\bibinfo {volume} {96}},\ \bibinfo {pages} {015003} (\bibinfo {year}
  {2006})}\BibitemShut {NoStop}%
\bibitem [{\citenamefont {Juan}\ and\ \citenamefont {I}(1998)}]{Juan}%
  \BibitemOpen
  \bibfield  {author} {\bibinfo {author} {\bibfnamefont {W.~T.}\ \bibnamefont
  {Juan}}\ and\ \bibinfo {author} {\bibfnamefont {L.}~\bibnamefont {I}},\
  }\href@noop {} {\bibfield  {journal} {\bibinfo  {journal} {Phys. Rev. Lett.}\
  }\textbf {\bibinfo {volume} {80}},\ \bibinfo {pages} {3073} (\bibinfo {year}
  {1998})}\BibitemShut {NoStop}%
\bibitem [{\citenamefont {Fortov}\ and\ \citenamefont
  {Morfill}(2010)}]{Fortov2010}%
  \BibitemOpen
  \bibfield  {author} {\bibinfo {author} {\bibfnamefont {V.~E.}\ \bibnamefont
  {Fortov}}\ and\ \bibinfo {author} {\bibfnamefont {G.~E.}\ \bibnamefont
  {Morfill}},\ }\href@noop {} {\emph {\bibinfo {title} {Complex and Dusty
  Plasmas, From Laboratory to Space}}}\ (\bibinfo  {publisher} {CRC Press, Boca
  Raton},\ \bibinfo {year} {2010})\BibitemShut {NoStop}%
\bibitem [{\citenamefont {Khrapak}\ and\ \citenamefont
  {Morfill}(2002)}]{Khrapak2}%
  \BibitemOpen
  \bibfield  {author} {\bibinfo {author} {\bibfnamefont {S.~A.}\ \bibnamefont
  {Khrapak}}\ and\ \bibinfo {author} {\bibfnamefont {G.~E.}\ \bibnamefont
  {Morfill}},\ }\href@noop {} {\bibfield  {journal} {\bibinfo  {journal} {Phys.
  Plasmas}\ }\textbf {\bibinfo {volume} {9}},\ \bibinfo {pages} {619} (\bibinfo
  {year} {2002})}\BibitemShut {NoStop}%
\bibitem [{\citenamefont {Ridolfi}, \citenamefont {D'Odorico},\ and\
  \citenamefont {Laio}(2011)}]{Ridolfi}%
  \BibitemOpen
  \bibfield  {author} {\bibinfo {author} {\bibfnamefont {L.}~\bibnamefont
  {Ridolfi}}, \bibinfo {author} {\bibfnamefont {P.}~\bibnamefont {D'Odorico}},
  \ and\ \bibinfo {author} {\bibfnamefont {F.}~\bibnamefont {Laio}},\
  }\href@noop {} {\emph {\bibinfo {title} {Noise-Induced Phenomena in the
  Environmental Sciences}}}\ (\bibinfo  {publisher} {Cambridge University
  Press, New York},\ \bibinfo {year} {2011})\BibitemShut {NoStop}%
\bibitem [{\citenamefont {Mori}(1965)}]{Mori}%
  \BibitemOpen
  \bibfield  {author} {\bibinfo {author} {\bibfnamefont {H.}~\bibnamefont
  {Mori}},\ }\href@noop {} {\bibfield  {journal} {\bibinfo  {journal} {Prog.
  Theor. Phys.}\ }\textbf {\bibinfo {volume} {33}},\ \bibinfo {pages} {423}
  (\bibinfo {year} {1965})}\BibitemShut {NoStop}%
\bibitem [{\citenamefont {Fortov}\ \emph {et~al.}(2004)\citenamefont {Fortov},
  \citenamefont {Khrapak}, \citenamefont {Khrapak}, \citenamefont {Molotkov},\
  and\ \citenamefont {Petrov}}]{Fortov2004}%
  \BibitemOpen
  \bibfield  {author} {\bibinfo {author} {\bibfnamefont {V.~E.}\ \bibnamefont
  {Fortov}}, \bibinfo {author} {\bibfnamefont {A.~G.}\ \bibnamefont {Khrapak}},
  \bibinfo {author} {\bibfnamefont {S.~A.}\ \bibnamefont {Khrapak}}, \bibinfo
  {author} {\bibfnamefont {V.~I.}\ \bibnamefont {Molotkov}}, \ and\ \bibinfo
  {author} {\bibfnamefont {O.~F.}\ \bibnamefont {Petrov}},\ }\href@noop {}
  {\bibfield  {journal} {\bibinfo  {journal} {Phys. Usp.}\ }\textbf {\bibinfo
  {volume} {47}},\ \bibinfo {pages} {447} (\bibinfo {year} {2004})}\BibitemShut
  {NoStop}%
\bibitem [{\citenamefont {Epstein}(1924)}]{Epstein}%
  \BibitemOpen
  \bibfield  {author} {\bibinfo {author} {\bibfnamefont {P.~S.}\ \bibnamefont
  {Epstein}},\ }\href@noop {} {\bibfield  {journal} {\bibinfo  {journal} {Phys.
  Rev.}\ }\textbf {\bibinfo {volume} {23}},\ \bibinfo {pages} {710} (\bibinfo
  {year} {1924})}\BibitemShut {NoStop}%
\bibitem [{\citenamefont {Baines}, \citenamefont {Williams},\ and\
  \citenamefont {Asebiomo}(1965)}]{Baines}%
  \BibitemOpen
  \bibfield  {author} {\bibinfo {author} {\bibfnamefont {M.~J.}\ \bibnamefont
  {Baines}}, \bibinfo {author} {\bibfnamefont {I.~P.}\ \bibnamefont
  {Williams}}, \ and\ \bibinfo {author} {\bibfnamefont {A.~S.}\ \bibnamefont
  {Asebiomo}},\ }\href@noop {} {\bibfield  {journal} {\bibinfo  {journal} {Mon.
  Not. R. Astron. Soc.}\ }\textbf {\bibinfo {volume} {130}},\ \bibinfo {pages}
  {63} (\bibinfo {year} {1965})}\BibitemShut {NoStop}%
\bibitem [{\citenamefont {Burov}\ and\ \citenamefont {Barkai}(2008)}]{Burov}%
  \BibitemOpen
  \bibfield  {author} {\bibinfo {author} {\bibfnamefont {S.}~\bibnamefont
  {Burov}}\ and\ \bibinfo {author} {\bibfnamefont {E.}~\bibnamefont {Barkai}},\
  }\href@noop {} {\bibfield  {journal} {\bibinfo  {journal} {Phys. Rev. Lett.}\
  }\textbf {\bibinfo {volume} {100}},\ \bibinfo {pages} {070601} (\bibinfo
  {year} {2008})}\BibitemShut {NoStop}%
\bibitem [{\citenamefont {Gel'fand}\ and\ \citenamefont
  {Shilov}(1964)}]{Gel'fand}%
  \BibitemOpen
  \bibfield  {author} {\bibinfo {author} {\bibfnamefont {I.~M.}\ \bibnamefont
  {Gel'fand}}\ and\ \bibinfo {author} {\bibfnamefont {G.~E.}\ \bibnamefont
  {Shilov}},\ }\href@noop {} {\emph {\bibinfo {title} {Generalized Functions,
  Properties and Operations}}}\ (\bibinfo  {publisher} {Academic Press, New
  York},\ \bibinfo {year} {1964})\BibitemShut {NoStop}%
\bibitem [{\citenamefont {Caputo}(1967)}]{Caputo}%
  \BibitemOpen
  \bibfield  {author} {\bibinfo {author} {\bibfnamefont {M.}~\bibnamefont
  {Caputo}},\ }\href@noop {} {\bibfield  {journal} {\bibinfo  {journal}
  {Geophys. J. R. Astr. Soc.}\ }\textbf {\bibinfo {volume} {13}},\ \bibinfo
  {pages} {529} (\bibinfo {year} {1967})}\BibitemShut {NoStop}%
\bibitem [{\citenamefont {Miller}\ and\ \citenamefont {Ross}(1993)}]{Miller}%
  \BibitemOpen
  \bibfield  {author} {\bibinfo {author} {\bibfnamefont {K.~S.}\ \bibnamefont
  {Miller}}\ and\ \bibinfo {author} {\bibfnamefont {B.}~\bibnamefont {Ross}},\
  }\href@noop {} {\emph {\bibinfo {title} {An Introduction to the Fractional
  Calculus and Fractional Differential Equations}}}\ (\bibinfo  {publisher}
  {John Wiley $\&$ Sons, New York},\ \bibinfo {year} {1993})\BibitemShut
  {NoStop}%
\bibitem [{\citenamefont {Podlubny}(1999)}]{Podlubny}%
  \BibitemOpen
  \bibfield  {author} {\bibinfo {author} {\bibfnamefont {I.}~\bibnamefont
  {Podlubny}},\ }\href@noop {} {\emph {\bibinfo {title} {Fractional
  Differential Equations}}}\ (\bibinfo  {publisher} {Academic Press, San
  Diego},\ \bibinfo {year} {1999})\BibitemShut {NoStop}%
\bibitem [{\citenamefont {Yates}\ and\ \citenamefont {Goodman}(2005)}]{Yates}%
  \BibitemOpen
  \bibfield  {author} {\bibinfo {author} {\bibfnamefont {R.~D.}\ \bibnamefont
  {Yates}}\ and\ \bibinfo {author} {\bibfnamefont {D.~J.}\ \bibnamefont
  {Goodman}},\ }\href@noop {} {\emph {\bibinfo {title} {Probability and
  Stochastic Processes: A Friendly Introduction for Electrical and Computer
  Engineers}}}\ (\bibinfo  {publisher} {John Wiley $\&$ Sons, New York},\
  \bibinfo {year} {2005})\BibitemShut {NoStop}%
\bibitem [{\citenamefont {Erd{\'{e}}lyi}(1981)}]{Erdelyi}%
  \BibitemOpen
  \bibfield  {author} {\bibinfo {author} {\bibfnamefont {A.}~\bibnamefont
  {Erd{\'{e}}lyi}},\ }\href@noop {} {\emph {\bibinfo {title} {Higher
  Transcendental Functions}}}\ (\bibinfo  {publisher} {Krieger, Malabar},\
  \bibinfo {year} {1981})\BibitemShut {NoStop}%
\bibitem [{\citenamefont {Haubold}, \citenamefont {Mathai},\ and\ \citenamefont
  {Saxena}(2011)}]{Haubold}%
  \BibitemOpen
  \bibfield  {author} {\bibinfo {author} {\bibfnamefont {H.~J.}\ \bibnamefont
  {Haubold}}, \bibinfo {author} {\bibfnamefont {A.~M.}\ \bibnamefont {Mathai}},
  \ and\ \bibinfo {author} {\bibfnamefont {R.~K.}\ \bibnamefont {Saxena}},\
  }\href@noop {} {\bibfield  {journal} {\bibinfo  {journal} {J. Appl. Math}\
  }\textbf {\bibinfo {volume} {2011}},\ \bibinfo {pages} {298628} (\bibinfo
  {year} {2011})}\BibitemShut {NoStop}%
\bibitem [{\citenamefont {Shukla}\ and\ \citenamefont
  {Prajapati}(2007)}]{Shukla}%
  \BibitemOpen
  \bibfield  {author} {\bibinfo {author} {\bibfnamefont {A.~K.}\ \bibnamefont
  {Shukla}}\ and\ \bibinfo {author} {\bibfnamefont {J.~C.}\ \bibnamefont
  {Prajapati}},\ }\href@noop {} {\bibfield  {journal} {\bibinfo  {journal} {J.
  Math. Anal. Appl}\ }\textbf {\bibinfo {volume} {336}},\ \bibinfo {pages}
  {797} (\bibinfo {year} {2007})}\BibitemShut {NoStop}%
\bibitem [{\citenamefont {Vi{\~{n}}ales}\ and\ \citenamefont
  {Desp{\'{o}}sito}(2006)}]{Vinales}%
  \BibitemOpen
  \bibfield  {author} {\bibinfo {author} {\bibfnamefont {A.~D.}\ \bibnamefont
  {Vi{\~{n}}ales}}\ and\ \bibinfo {author} {\bibfnamefont {M.~A.}\ \bibnamefont
  {Desp{\'{o}}sito}},\ }\href@noop {} {\bibfield  {journal} {\bibinfo
  {journal} {Phys. Rev. E}\ }\textbf {\bibinfo {volume} {73}},\ \bibinfo
  {pages} {016111} (\bibinfo {year} {2006})}\BibitemShut {NoStop}%
\bibitem [{\citenamefont {Pottier}(2003)}]{Pottier2003}%
  \BibitemOpen
  \bibfield  {author} {\bibinfo {author} {\bibfnamefont {N.}~\bibnamefont
  {Pottier}},\ }\href@noop {} {\bibfield  {journal} {\bibinfo  {journal}
  {Physica A}\ }\textbf {\bibinfo {volume} {317}},\ \bibinfo {pages} {371}
  (\bibinfo {year} {2003})}\BibitemShut {NoStop}%
\bibitem [{\citenamefont {Kneller}(2011)}]{Kneller}%
  \BibitemOpen
  \bibfield  {author} {\bibinfo {author} {\bibfnamefont {G.~R.}\ \bibnamefont
  {Kneller}},\ }\href@noop {} {\bibfield  {journal} {\bibinfo  {journal} {J.
  Chem. Phys.}\ }\textbf {\bibinfo {volume} {134}},\ \bibinfo {pages} {224106}
  (\bibinfo {year} {2011})}\BibitemShut {NoStop}%
\bibitem [{\citenamefont {Green}(1954)}]{Green}%
  \BibitemOpen
  \bibfield  {author} {\bibinfo {author} {\bibfnamefont {M.~S.}\ \bibnamefont
  {Green}},\ }\href@noop {} {\bibfield  {journal} {\bibinfo  {journal} {J.
  Chem. Phys.}\ }\textbf {\bibinfo {volume} {22}},\ \bibinfo {pages} {398}
  (\bibinfo {year} {1954})}\BibitemShut {NoStop}%
\bibitem [{\citenamefont {Kubo}(1957)}]{Kubo}%
  \BibitemOpen
  \bibfield  {author} {\bibinfo {author} {\bibfnamefont {R.}~\bibnamefont
  {Kubo}},\ }\href@noop {} {\bibfield  {journal} {\bibinfo  {journal} {J. Phys.
  Soc. Jpn.}\ }\textbf {\bibinfo {volume} {12}},\ \bibinfo {pages} {570}
  (\bibinfo {year} {1957})}\BibitemShut {NoStop}%
\end{thebibliography}%

\end{document}